\documentclass[preprint,showpacs,preprintnumbers,amsmath,amssymb]{revtex4}
\usepackage{epsfig}
\usepackage{graphicx}
\usepackage{dcolumn}
\usepackage{bm}

\begin{document}

\title{Island coarsening in one-dimensional models with partially and completely
reversible aggregation}
\author{Anna Chame }  
\author{F. D. A. Aar\~ao Reis}          
\affiliation{%
Instituto de F\'\i sica, Universidade Federal Fluminense,\\
Avenida Litor\^anea s/n, 24210-346 Niter\'oi RJ, Brazil}
\date{\today}

\begin{abstract}

Using computer simulations and scaling ideas, we study one-dimensional models of
diffusion, aggregation and detachment of particles from islands in the
post-deposition regime, i. e. without flux. The diffusion of isolated particles
takes place with unit rate, aggregation occurs immediately upon contact with
another particle or island, and
detachment from an island occurs with rate $\epsilon = exp(-E/kT)$, where $E$
is the related energy barrier. In the partially reversible model, dissociation
is
limited to islands of size larger than a critical value $i$, while in the
completely reversible model there is no restriction to that process (infinite
$i$). Extending previous simulation results for the completely reversible case,
we observe that a peaked island size distribution in the intermediate time
regime, in which the mean island size is increasing, crosses over to the
theoretically predicted exponentially decreasing distribution at long times. It
contrasts with the partially
reversible model, in which peaked distributions are obtained until the long
time frozen state, which is attained with a
crossover time $\tau \sim \frac{i^3}{\epsilon}$. The mean island size at
saturation varies as $S_{sat}\approx 2i+C\epsilon$ ($C$ constant), while the
completely reversible case shows an Ahrrenius dependence of the mean island
size, $S\sim \epsilon^{-1/2}$. Thus,
for different coverages, the effect of the critical size $i$ on the geometric
features is much stronger than that of $\epsilon$, which may be used to infer
the relevance of size-dependent detachment rates in real systems and other
models.

\end{abstract}

\pacs{ 68.43.Jk,  81.10.Aj, 68.55.Ac}  

\maketitle

\section{Introduction}

The early stages of formation of a thin film (submonolayer regime)
is dominated by the deposition and diffusion of adatoms and the formation and
growth of  islands, through nucleation,  coalescence and capture of new
adatoms. This  problem  has long been of fundamental and technological interest
 \cite{Venables} because those processes also have  a strong influence on the
later stages of growth. More recently, this interest increased for the
possibility of formation of  novel nano-structures.

Systems where diffusion-mediated nucleation of islands competes 
with deposition of particles have been extensively studied
both experimentally and theoretically \cite{Evans2006,Metois79,Brune98}. From
the
theoretical point of view, island formation (in one and two dimensions) has
mainly been treated using scaling theories, rate equations and simulations.
The models of deposition,
diffusion and aggregation may be separated in three groups according to the
conditions for the islands stability: models with completely irreversible
aggregation \cite{BarteltEvans96,AmarFamilyLam,Popescu01}, with completely
reversible aggregation \cite{Ratsch94,BalesZangwill97},  or those which assume
the existence of a  critical island size $i$, above which
islands are stable against dissociation \cite{Venables,AmarFamily95,Kandel97}.
The critical island size is usually expected to
represent thermal effects and the substrate geometry; for instance, $i=2$ 
corresponds to stable trimers in a triangular lattice and $i=3$ to stable
tetramers in a square lattice \cite{AmarFamily95}. Other mechanisms that make
island growth to be size-dependent may also justify such conditions.

Another important problem is the nucleation and growth of islands in the
post-deposition regime, i. e. after the particle flux has stopped. 
Diffusion and nucleation in this regime are observed in several experiments
~\cite{Rosenfeld93,Mueller96,Evans2003}.  
One possible situation is growth at low enough temperatures so that adatom 
diffusion is negligible during deposition, but with
adatom rearrangement into islands occurring during a much longer time interval
after the flux has been shut off \cite{Mueller96,Evans2003}. This is certainly
relevant for technological applications in which a system needs to have its
properties preserved for long times. Other possible situation is that in which
the sample is annealed ~\cite{Himpsel2001}. For the case of completely
irreversible aggregation ($i=1$), the post-deposition regime was studied in one
and two dimensions by several authors \cite{Lam99,Tataru00,Ammi}. The case
$i=3$ was also considered in two-dimensions by Tataru et al \cite{Tataru00}.
Only for a point island model ~\cite{Li1997}, the case of
partially reversible aggregation in a one-dimensional substrate has  been 
treated in the post-deposition regime for several values of the critical size
($i=1$, $ 2$ and $3$). On the other hand, but also in one dimension, this
post-deposition  regime has been studied for completely reversible 
aggregation with an extended island model \cite{ReisSt2004}.

In this work, we consider one-dimensional models with critical size $i$ for 
islands coarsening in the post-deposition regime, focusing on the differences
from the completely reversible situation ($i=\infty$), whose results from
previous works are extended. First, the particles are
randomly deposited on a one-dimensional lattice, and diffusion, aggregation
and detachment mechanisms subsequently take place. Diffusion of isolated
particles occurs with rate $D=1$ (probability $1/2$ to move to each neighboring
empty site), aggregation of a free particle immediately takes place upon
contact with another particle or cluster, and detachment occurs with rate
$\epsilon$ if the cluster size does not exceed the critical size $i$ (Fig. 1b).
The unit rate for hopping of isolated particles defines the time scale
of the problem, and the detachment rate $\epsilon$ is of order
$exp(-E/k_BT)$, where E is an energy barrier for dissociation.
The time evolution of island size will be analyzed for different
detachment rates, coverages and critical sizes, and the island size
distributions in different regimes will be obtained.
One of our aims is to know how suitable is the
approximation  of finite critical size to describe the island coarsening
without particle flux in different time regimes.

At first sight, the restricted dimensionality considered here seem to have
limited interest, but there are experimentally important cases of aggregate
confinement, such as nucleation processes occurring along surface steps
\cite{Himpsel2001,Gambardella2000,Gai2002}. 
Moreover, some experiments show atoms diffusing along steps 
without detaching from them at room temperature \cite{Giesen2001}, the formation
of one-dimensional nano-structures on patterned substrates \cite{Nguyen,Notzel}
and the formation of single-atom-wide metal rows due to inhibition of
aggregation at their sides \cite{albao}. Consequently,
exploring nontrivial differences between these relatively simple models may be
important for comparison with experimental data, as well as to design more
complex models. This is illustrated, for instance, in recent work on
two-dimensional irreversible island growth \cite{politi}. For these reasons, we
will perform a careful quantitative analysis of the island properties and
the time scales in the models presented here.

The rest of this work is organized as follows. In Sec. 2 we present
a brief review of previous results for the fully reversible model and simulation
data for the island size distribution, whose shape shows significant changes in
time. In Sec. 3 we discuss our results for the
model with partially reversible aggregation. In Sec. 4 we summarize our
results and present our conclusions.

\begin{figure}[!ht]
\centering
\includegraphics[clip,width=0.95\textwidth,
height=0.10\textheight,angle=0]{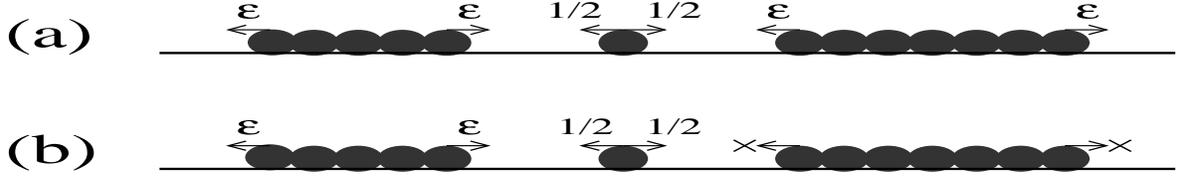}
\caption{
Diffusion and detachment processes of (a) the completely reversible model
and (b) the model with a critical size. Detachment from islands with sizes
greater than $i$ is forbidden (denoted by x).
}
\label{model}
\end{figure}

\section{The completely reversible model}

For the completely reversible model in the post-deposition 
regime \cite{ReisSt2004}, the particles are randomly deposited on a 
one-dimensional substrate at $t=0$ and immediately begin to diffuse according to
the conditions of Fig. 1a, but with no restriction to detach from an
island ($i=\infty$).

The coarsening in this system was shown to be separated in three main regimes
\cite{ReisSt2004}: (i) early fast attachment of isolated particles to each
other;
(ii) an intermediate regime in which detachment sets in, 
the mean island size increasing as
\begin{equation}
S  \sim {(\epsilon t)}^{1/3}
\label{sfd}
\end{equation}
in the limit  of small $\epsilon$; (iii) a slow (diffusive) evolution of the
average cluster
size as
\begin{equation}
S(t) = S_{\infty} -C/t^{1/2} ,
\label{scrosfd}
\end{equation}
where
\begin{equation}
S_{\infty} \sim \epsilon^{-1/2}
\label{sinf}
\end{equation}
for small $\epsilon$. 
This last regime is diffusion-limited 
in the sense that the distances between clusters are still large so that a  
detached particle probably returns and reattaches 
many times before effectively diffuse to the nearest cluster.
Actually this system does not attain neither an equilibrium nor a steady state.
The behavior in regime (iii) and the corresponding island size distribution were
predicted by a solution of the master equation in an independent interval
approximation, while the behevior in regime (ii) was derived using scaling
arguments \cite{ReisSt2004}.

Extending previous kinetic Monte Carlo simulation,
we confirmed the theoretically predicted island size distribution in the
asymptotic state [regime (iii)] \cite{ReisSt2004}:
\begin{equation}
P(x)\sim \epsilon \exp{\left( -\frac{x}{S}\right)} ,
\label{distrinf}
\end{equation}
where $x$ is the island size and $S\approx \epsilon^{-1/2}
{\frac{\theta}{2\left( 1-\theta\right)}}^{1/2}$ is the average size for coverage
$\theta$. This exponentially decreasing distribution is shown in Fig. 2  for
$\epsilon= 1/128$ and $\theta = 0.5$, at a sufficiently long time 
$t= 2 \times 10^{5}$ where the
average cluster size is $S=9.53 $ ($S_\infty = 9.99 $ in this case).
However, the distribution at early times [regime (ii)] shows a peak at a certain
typical size $S\approx 4 $, as illustrated in Fig. 2 for the same model
parameters and $t= 3000$. This new result was not predicted by the analytical
solution which explained regime (iii). On the other hand, it is probably the
reason for the successful derivation of Eq. (\ref{sfd}) with a scaling approach
that assumed the existence of a typical cluster size in the system in regime
(ii). This assumption fails in the case of a very large number of small
clusters, such as the asymptotic limit (iii).

\begin{figure}[!ht]
\centering
\includegraphics[clip,width=0.95\textwidth,
height=0.30\textheight,angle=0]{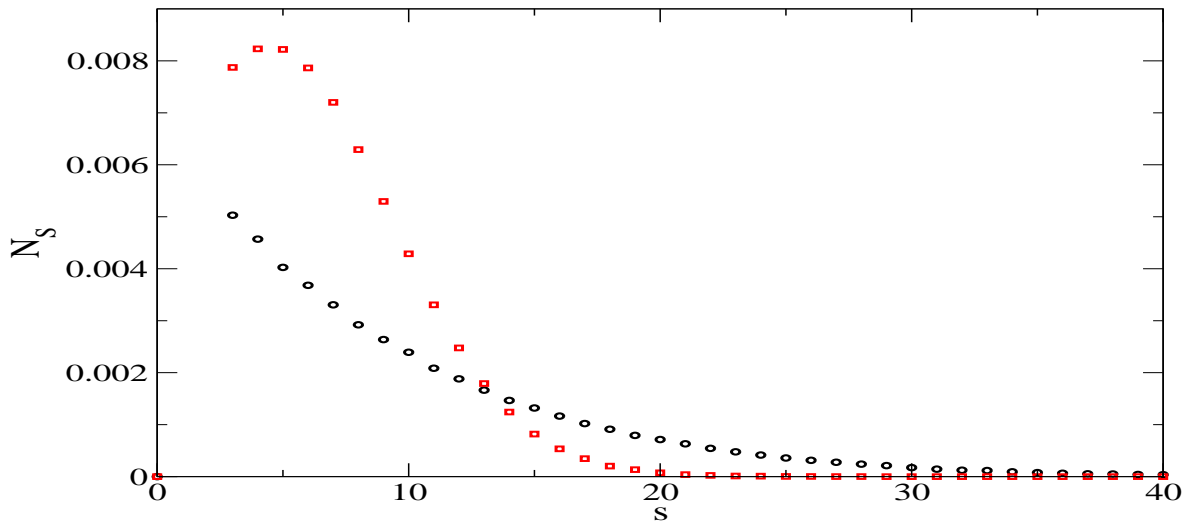}
\caption{Island size distributions for the completely reversible model,
for $\epsilon = 1/128$, $\theta = 0.5$ and $L= 8000$.
At  time $t = 3000$, the distribution presents a peak (squares), and at
time $t = 2 \times 10^5$ the distribution decreases monotonically (circles).
Averages over 1000 realizations.
}
\label{dist}
\end{figure}

The crossover from regime (ii) to regime (iii) is expected to take place at
times of the order $\epsilon^{-5/2}$, which can be obtained by matching the
typical sizes in Eqs. (\ref{sfd}) and (\ref{sinf}). The above results show that
this is the expected time for a crossover from a peaked island size
distribution to a monotonically decreasing one. This difference may be
important for comparisons with experimental data and, as will be shown below,
for a test of the reliability of the assumption of a critical island size.

At this point, the comparison with results for reversible models where
deposition competes with
aggregation is also important. Without accounting for the post-deposition
behavior, peaked distributions are observed in several
conditions \cite{Ratsch94}. Moreover, they were also obtained by Lam and
co-workers \cite{Lam99} in the post-deposition regime, although a
systematic search for an asymptotic regime was not performed there. Since this
type of distribution is obtained in regime (ii) of our model, one
naturally raises the question whether such crossover is also present in two
dimensions. On the other hand, monotonically decreasing size distributions of
one-dimensional islands were obtained in Ref. \protect\cite{albao}, and were
explained by a highly anisotropic model of diffusion and completely
irreversible aggregation ($i=1$) in a square lattice.

In models with conserved density, it is interesting to mention the case of Ref.
\protect\cite{girardi}, in which increasing the concentration of amphiphiles
leads to a crossover from monotonically decreasing cluster size distributions
to peaked ones. Moreover, depending on the particular spin dynamics, both types
of distributions were found with same temperature and concentration.

\section{ The partially reversible model }

In this model, after random deposition of a coverage $\theta$ at $t=0$, 
diffusion, detachment and  re-attachment of particles
continue until all islands are stable (size larger than $i$), following the
rules of Fig. 1b. We simulated this model in lattices 
of length $L= 8000$, which are large enough to avoid finite-size 
effects, with coverages $\theta = 0.1$, $0.5$ and $ 0.9$. The critical sizes
analyzed here range from $2$ to $70$ and the
diffusion rates $\epsilon$ range from $10^{-1}$ to $10^{-4}$.

We observed a sequence of  four regimes of aggregation in
the simulations: (a) fast attachment of isolated particles to each other to form
islands (b) an intermediate regime in which detachment sets
in, allowing further coarsening, but the size of the
islands are typically below $i$, thus with negligible influence of the  critical
size; (c) a crossover region, which begins when a large quantity of
stable islands are formed; (d) a frozen state where
there is no more isolated atoms nor unstable islands.
Regimes (a) and (b) are equivalent to regimes (i) and (ii) of the completely
reversible model, which is justified by simulation results.

In all cases, the mean size $S$ increases in time and attains
a saturated value $S_{sat}$ that increases with $i$, for fixed $\epsilon$ and
$\theta$. This is illustrated in Fig. 3a, where we compare the
time evolution of $S$ for $i=10$, $i=15$ and $i=50$, with $\epsilon = 10^{-4}$.
In Fig. 3b we compare the result for $i=50$ with that for the completely
reversible model with the same $\epsilon$.

The results of Fig. 3a show that small values of $i$ have drastic effects at
intermediate times. Due to the stability of small clusters, the mean size $S$
increases much earlier and exceeds the mean size for large values of $i$. For
instance, $S$ for $i=10$ is larger than $S$ for $i=50$ in a time interval which
extends for more than one decade in $\log{(t)}$. The very slow growth in regime
(b) [or (ii) in the completely reversible case] contributes to this effect.
Consequently, the short-time behavior of $S$ does not reveal the true
asymptotics ($S_{sat}$ increasing with $i$) in cases where the critical size is
a realistic assumption. 

These results are related to the fact that
the intermediate regime (b) is clear only for relatively large $i$, otherwise
the crossover to saturation takes place immediately after formation of small
islands. Moreover, an increase of the mean cluster size similar to the fully
reversible model [Eq. \ref{sfd}, regime (ii)] is observed only for very small
$\epsilon$, typically $\epsilon\leq 10^{-4}$. This illustrated in Fig. 3b, where
we show a line with slope $1/3$, which is nearly parallel to the data for both
models in that regime.

\begin{figure}[!ht]
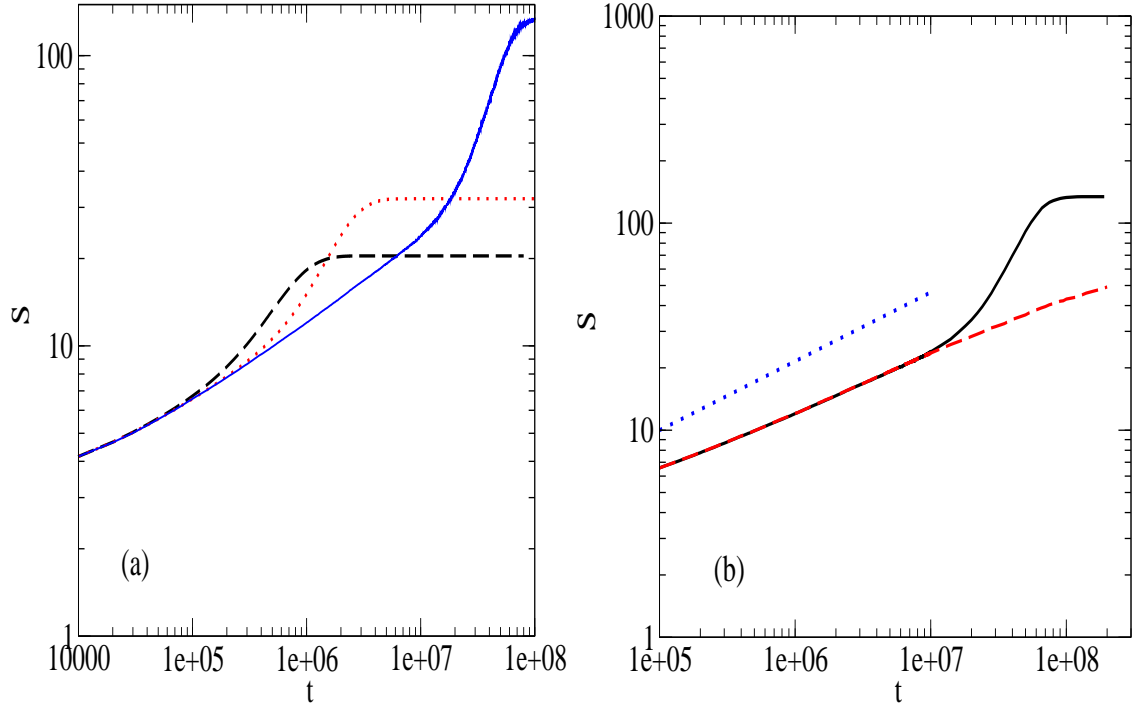

\includegraphics[clip,width=0.45\textwidth,
height=0.40\textheight,angle=0]{fig3a.eps}
\includegraphics[clip,width=0.45\textwidth,
height=0.40\textheight,angle=0]{fig3b.eps}
\caption{ 
Time evolution of the mean island size S, for $\theta = 0.5$, $\epsilon =
0.0001$ and $L=8000$ (a) for different values of the critical size:
$i=10$ (dashed) , $i=15$ (dotted),  averages  over 1000 realizations,
and $i=50$ (solid), averages over 50 realizations; (b)for $i= 50$ (solid) and
for the fully reversible aggregation model (dashed). The dotted line has slope
$1/3$.
}
\label{s_of_i}
\end{figure}

Contrary to the long time diffusive behavior of the completely reversible model
(Eq. \ref{scrosfd}), in the present case the crossover to the frozen state,
regime (c), shows the expected exponential decay to a saturation value, as
\begin{equation}
S(t) = S_{sat} - e^{(-t/\tau)} .
\label{ssat}
\end{equation}
This is illustrated in Fig. 4 for $i=5$, $\epsilon= 10^{-2}$ and $\theta= 0.5$.
Smaller values of $i$ show much longer and well defined convergences to the
saturation sizes.

For fixed $i$, the mean island size in the frozen state, $S_{sat}$, 
increases approximately as
\begin{equation}
S_{sat} = S_0 + C \epsilon ,
\label{ssat1}
\end{equation}
typically with small $C <  100$, which corresponds to a very weak 
dependence on
$\epsilon$ in the low temperature regime. This is illustrated in Fig. 5, where
we show $S$ versus $\epsilon$ for
$i= 5$ and different values of the coverage $\theta$. Except for coverages very
near $1$, the saturation value for
low temperatures ($\epsilon\ll 1$) is nearly $S_0\approx 2i$; for instance,
$S_0\approx 10$ in Fig. 5 for $\theta=0.1$ and $\theta=0.5$. This expected
proportionality between $S_0$ and $i$ is also observed for larger values of i.

\begin{figure}[!ht]
\centering
\includegraphics[clip,width=0.95\textwidth,
height=0.40\textheight,angle=0]{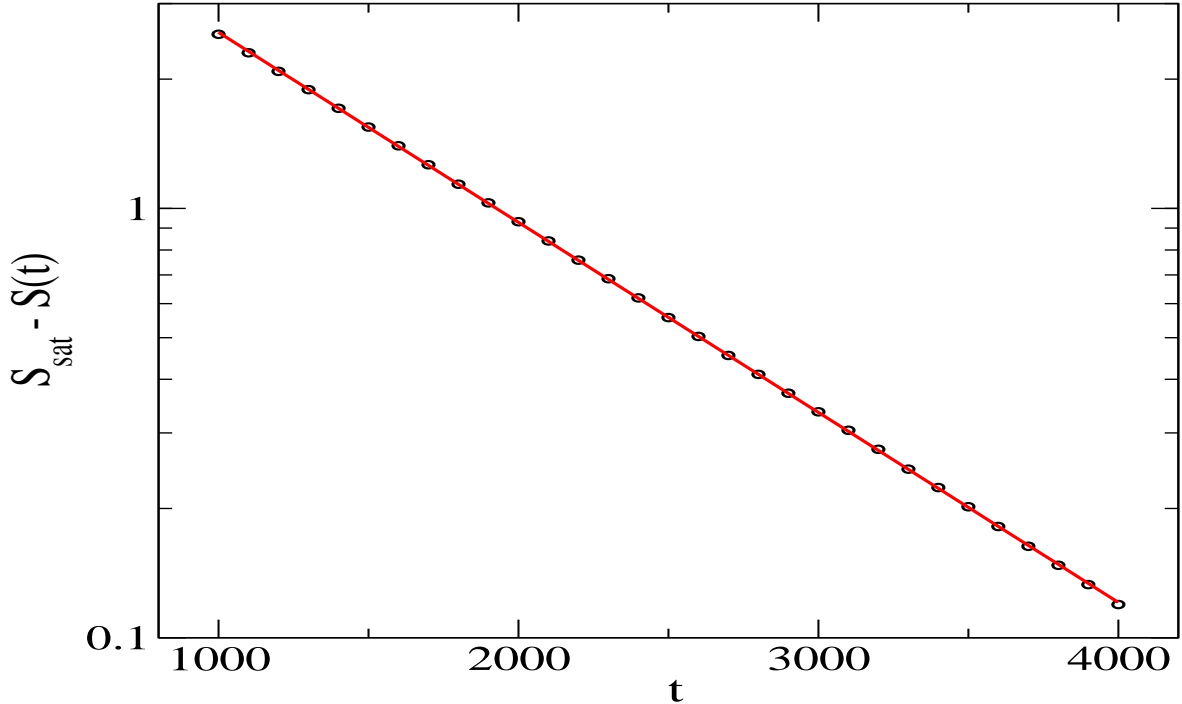}
\caption{$S_{sat} - S(t)$ for $i=5$, $\theta = 0.5$, $\epsilon= 10^{-2}$, $L = 8000$,
averaged over 1000 realizations.  $S_{sat}-S(t)= 7.12 e^{(-t/981)}$
}
\label{fig:exp}
\end{figure}

\begin{figure}[!ht]
\centering
\includegraphics[clip,width=0.95\textwidth,
height=0.40\textheight,angle=0]{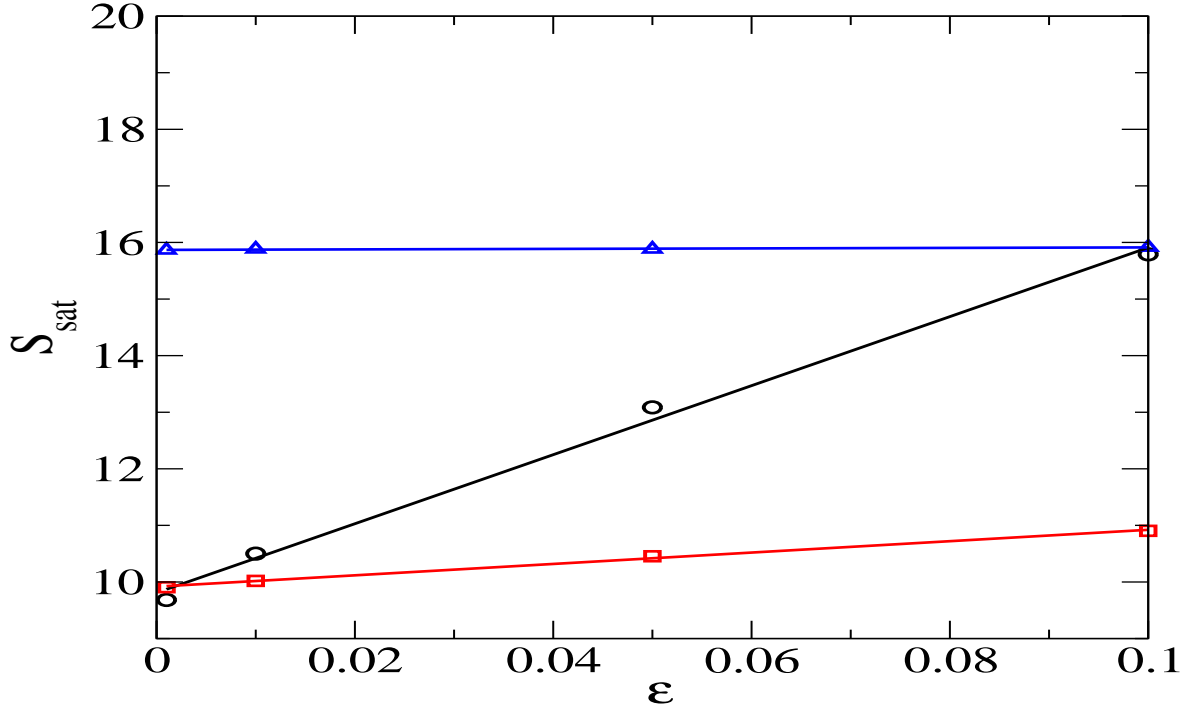}
\caption{Mean island size at the
final state, $S_{sat}$, as a function of  $\epsilon$, for $i=5$, $\theta= 0.1 $
(circles), $\theta = 0.5$ (squares), $\theta = 0.9$ (triangles).
}
\label{fig: ssat}
\end{figure}

This result contrasts to the remarkable dependence of the mean island size on
$\epsilon$ for the completely reversible case [regime (iii), Eq. \ref{sinf}].
Here, the critical island size is the main parameter to determine the features
of the system in the saturation regime, while $\epsilon$ has an important role
in determining the time scale of the problem. It is also the parameter $i$ which
is the more closely connected to other geometric features, such as size
distributions (discussed below).

The crossover time $\tau$ in Eq. (\ref{ssat}) can be estimated from the
previously discussed features of both models. First recall that Eq. (\ref{sfd})
governs the time evolution of the mean island size in intermediate times when
both models have similar features (small $\epsilon$, large $i$). On the other
hand, at the crossover time $\tau$, this size must match the saturation value
$S\approx 2i$ for the present model ($S\sim i$ is certainly intuitive). Thus we
obtain
\begin{equation}
\tau \sim \frac{i^3}{\epsilon} .
\end{equation}
The result $\tau\sim 1/\epsilon$ is expected because the detachment rate defines
the time scale of the model; it is confirmed by simulation for small
$\epsilon$ with good accuracy. However, much more interesting is the nontrivial
dependence of $\tau$ on $i$. In Fig. 6, it is confirmed by simulation with
large values of $i$  ($i=30$ to $i=70$), for $\epsilon = 10^{-4} $ and $\theta
= 0.5$. Notice that, despite the small range of values of $i$, $i^3$ varies by
a factor larger than $5$ in Fig. 6. This scaling certainly requires relatively
large values of $i$ and small $\epsilon$ to be observed.
On the other hand, this analysis reinforces the conclusion that the intermediate
time regime of this model, regime (b), has negligible differences from the
model with completely reversible aggregation.

The final distribution of island sizes (frozen state) has a pronounced  peak
for small $i$, due to the stability of small clusters that stop feeding the
largest ones at early times. As expected, the distribution becomes
larger when $i$ increases, as illustrated in Fig. 7. For  fixed $i$ and
$\theta$, the distribution becomes narrower for smaller $\epsilon$.

\begin{figure}[!ht]
\centering
\includegraphics[clip,width=0.95\textwidth,
height=0.40\textheight,angle=0]{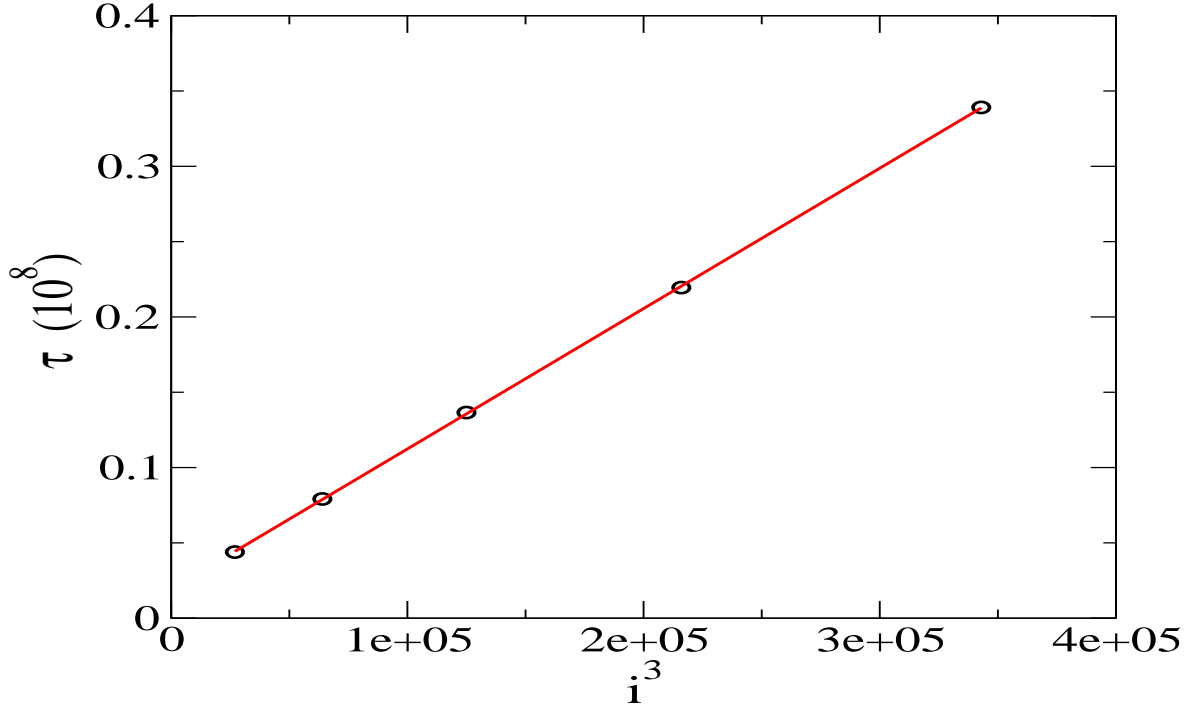}
\caption{
Dependence of the crossover time  $\tau$  on the
critical size $i$. For $i=30,40,50,60$ and $70$,
$\theta = 0.5$, $ L = 8000$, averaged over 50 realizations.
}
\label{fig:taui3}
\end{figure}

\begin{figure}[!ht]
\centering
\includegraphics[clip,width=0.95\textwidth,
height=0.40\textheight,angle=0]{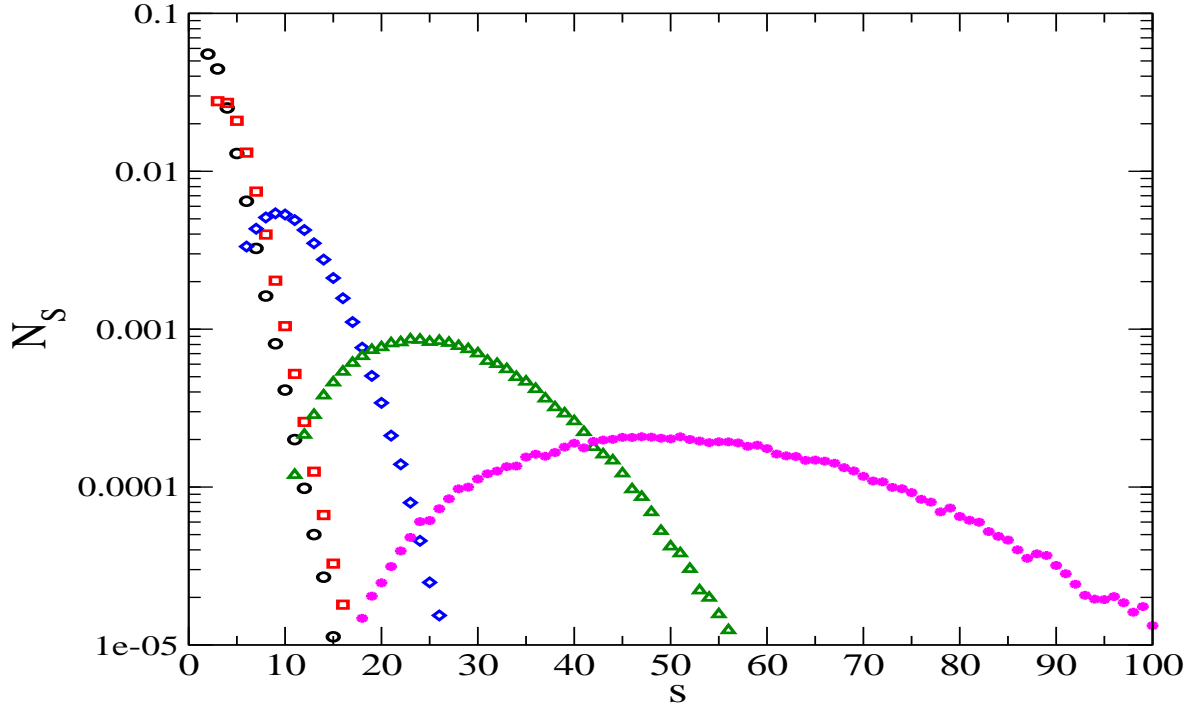}
\caption{
Island size distribution for  $\theta = 0.5$, $\epsilon = 0.1$
$i= 1$(circles), $2$ (squares), $5$ (diamonds), $10$(triangles), $15$ (stars).
}
\label{disti}
\end{figure}

In order to illustrate the remarkable difference from the fully reversible
model at long times, we compared the island size distributions for the 
same $\epsilon$ and $S_{sat}\approx S_{\infty}$, using $\theta= 0.5$. 
For instance, for $i=15$ and
very small $\epsilon$ we obtain $S_{sat}\approx 32$, and the same value for
$S_{\infty}$ is obtained in the fully reversible model with $\epsilon = 
0.00049$. The distributions of both models are compared in Fig. 8a, with the
same value of $\epsilon$ and $i=15$ for the model with critical size. The same
comparison is
also performed in Fig. 8b for $i=5$, $\theta = 0.1$, and $\epsilon = 0.000845$,
where $S_{sat}\approx 9.5$ in both models. Thus, from the shape of the size
distribution at very long times, we are able to test the reliability of each
model for a given system.

\begin{figure}[!ht]
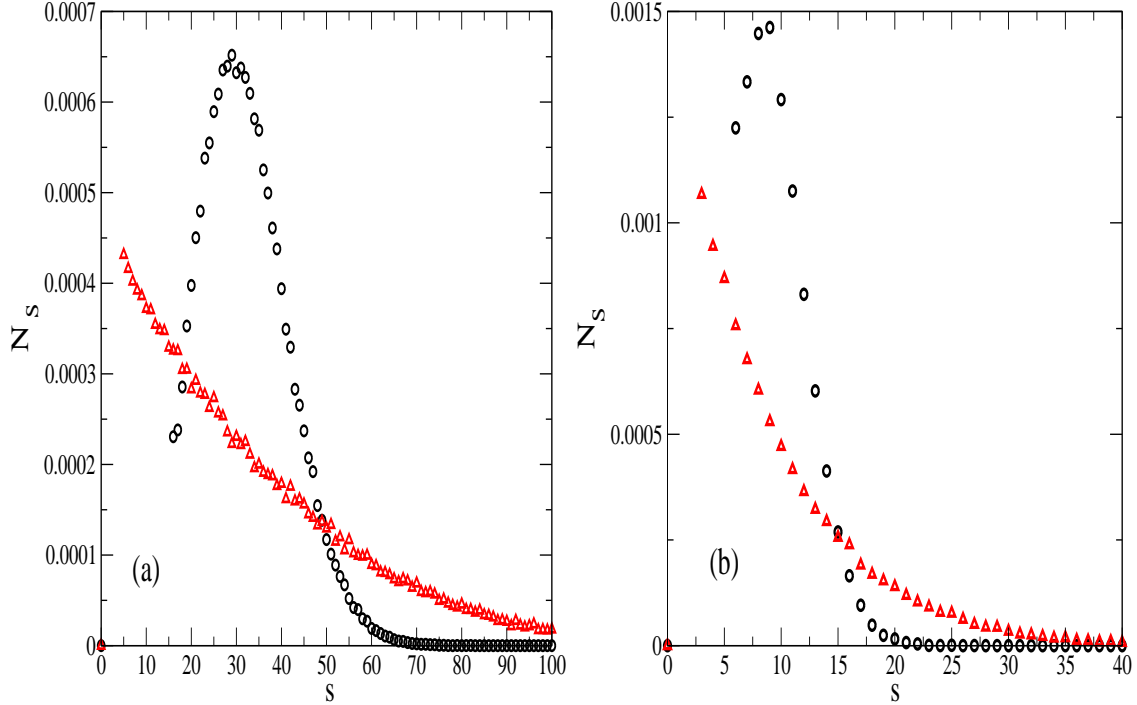

\includegraphics[clip,width=0.45\textwidth,
height=0.40\textheight,angle=0]{fig8a.eps}
\includegraphics[clip,width=0.45\textwidth,
height=0.40\textheight,angle=0]{fig8b.eps}
\caption{
Island size distributions for the same values of
$S_{sat}$ and $\epsilon$, using (a) $\theta= 0.5$ and
$\epsilon = 0.0049 $, $S_{sat}\approx 32$ , for $i=15$ (circles) at the
final state, and for the fully
reversible model (triangles), at $t= 5 \times 10^{7}$ (asymptotic regime).
(b) $\theta = 0.1$  and $\epsilon = 0.000845$, $S_{sat} \approx 9.5$,  for
$i=5$ (circles) at the final state and
for the totally reversible model (triangles), at $t = 3 \times 10^7  $
(asymptotic regime).
}
\label{fig:distr}
\end{figure}

\section{Conclusion}

We studied island coarsening in the post-deposition regime in one-dimensional
models with completely and with partially reversible aggregation, the latter
with critical size $i$. Extending previous work on the fully reversible case,
here we showed that the island size distributions are very different in the
regime
of island coarsening and in the long time limit, crossing over from a peaked
distribution to an exponentially decreasing one. For the model with critical
size $i$, we systematically studied the influence of this parameter, 
the detachment  rate   $\epsilon$ and the coverage $\theta$  on the mean island
size and the island size distribution .
Contrary to the long time diffusive increase of the average island size in the 
fully reversible model, for the partially reversible model the crossover to 
a frozen state shows an exponential decay to the saturation value, with
crossover time $ \tau \sim \frac{i^3}{\epsilon}$, a relation which follows from
the fact that, at intermediate times, both models show the same coarsening.
For fixed $i$, the mean island size in the frozen state
has a very weak dependence on $\epsilon$, in contrast to the completely
reversible model, where it varies as $S\sim \epsilon^{-1/2}$.
Another particularly interesting result is that the peaked island size
distribution at intermediate times for both models crosses over to a long time
exponentially decreasing distribution only for the completely reversible model.

We believe that some of these features may motivate comparisons with real
systems with effective one-dimensional behavior (see e. g. the recent discussion
in Ref. \protect\cite{phillips}. For instance, the completely reversible model
predicts an Ahrrenius
temperature dependence of the mean island size (Eq. \ref{sinf} with
$\epsilon\sim \exp{\left( -E/k_BT\right)}$) in the long time limit, i. e., when
that size does not show significant time increase. Deviation from this
dependence may suggest some size-dependence of detachment rates, and would
eventually justify the assumption of a critical island size or another
restriction to the growth of large islands for modeling the island coarsening.
Moreover, peaked island size distributions may indicate deviations from the long
time conditions of the completely reversible model, which may be interpreted as
a finite-time behavior [regime (ii) of the same model] or as an effect of
size-dependent detachment rates.

There are also many recent examples in which the solution of simple statistical
models in one-dimension anticipate features of the more realistic two- or
three-dimensional systems, such as non-equilibrium multilayer growth models
\cite{sukarno}, magnetic systems with rough surfaces \cite{noninteger} and
other aggregation models \cite{salazar}. Thus, further systematic studies of
island coarsening in the post-deposition regime in two dimensions are certainly
motivated by the present work.

\pagebreak

\end{document}